\documentclass[fp,dvipdfmx]{jpsj3}
\usepackage[dvipdfmx]{graphicx}
\usepackage[dvipdfmx]{color}
\renewcommand{\a}{\alpha}
\renewcommand{\b}{\beta}
\newcommand{\bea}{\begin{eqnarray}}
\newcommand{\eea}{\end{eqnarray}}
\newcommand{\f}[2]{\frac{#1}{#2}}
\newcommand{\eq}{&=&}
\newcommand{\nn}{\nonumber \\ }
\newcommand{\ve}{\varepsilon}

\newcommand{\area}{\int_{-\infty}^\infty }

\newcommand{\p}{\partial}
\newcommand{\pp}[2]{\f{\p #1}{\p #2}}

\newcommand{\siki}[1]{Eq. (\ref{#1})}
\newcommand{\sikis}[2]{Eqs. (\ref{#1}) and (\ref{#2})}

\newcommand{\kitai}[1]{\left\langle #1\right\rangle}
\allowdisplaybreaks[1]

\allowdisplaybreaks[1]

\title{
Macroscopic theorem of the portfolio optimization problem\\
with a risk-free asset
}

\author{Ippei Suzuki and Takashi Shinzato\thanks{Corresponding author: shinzato@eng.tamagawa.ac.jp}
}
\inst{
Department of Management Science, College of Engineering, 
Tamagawa University, \\ Machida, Tokyo 1948610, Japan\\
\smallskip
{\rm (Received June 21, 2019; accepted *** 1, 2019)}
} 
\abst{
The investment risk minimization problem with budget and return constraints has been the subject of research using replica analysis but there are shortcomings in the extant literature. 
With respect to  
 {Tobin's separation} theorem 
and the capital asset pricing model,
it is necessary to investigate the implications of a risk-free asset and examine its influence on
the optimal portfolio. Accordingly, in this work, 
{we explore the investment risk} minimization problem in the presence of a risk-free asset with {budget and return constraints.} Moreover, we discuss 
opportunity loss, the 
Pythagorean theorem of the Sharpe ratio, and Tobin's
separation theorem.
}
\begin{document}
\maketitle

\section{Introduction}
The problem of portfolio optimization, which is important from the perspective of asset management, 
has been discussed in 
the pioneering work reported by Markowitz and in various studies within the domain of operations research. In the framework of the stochastic optimization problem, it has been pointed out that, in recent years, 
the findings of 
this optimization problem by operations research is not an investment scenario that responds to the optimal investment strategy required by a rational investor\cite{
Ciliberti2007,
10.1371/journal.pone.0134968,
KONDOR20071545,
Pafka2002,PAFKA2003487,
doi:10.1080/14697680701422089,
doi:10.1080/1351847X.2011.601661,
110008689817,
1742-5468-2017-12-123402,
1742-5468-2016-12-123404,
doi:10.7566/JPSJ.87.064801,
2018arXiv181006366S,
10.1371/journal.pone.0133846,
PhysRevE.94.062102b,
1742-5468-2017-2-023301,
SHINZATO2018986,
doi:10.7566/JPSJ.86.124804,
Ryosuke-Wakai2014,
doi:10.7566/JPSJ.86.063802,
PhysRevE.94.052307,
1742-5468-2018-2-023401}.
In interdisciplinary research within the framework of stochastic optimization, for example, the following are examined using techniques such as replica analysis and the Thouless-Anderson-Palmer equation: finding the ground state of the quenched disorder system of the Sherrington-Kirkpatrick model in spin glass theory by the absolute zero limits; and/or finding the ground state that minimizes the Hamiltonian defined by embedded patterns in the associative memory problem. Thus, the importance of ground state analysis in quenched disordered systems is commonly recognized. Although the portfolio optimization problem is formulated in the framework of stochastic optimization, in conventional operations research, analysis has mainly been conducted with respect to the annealed disordered system in spin glass theory.
However, since the optimal investment strategy required by rational investors corresponds approximately to the ground state derived by the approach to the quenched disordered system in spin glass theory, in recent decades, research has been conducted to robustly evaluate the optimal solution of portfolio optimization problems using interdisciplinary analytical methods, for instance, replica analysis, the belief propagation method, and random matrix theory.

For instance, Ciliberti et al.\ explore  
the minimal investment risk of the absolute deviation model and/or the expected shortfall model described by a perceptron-type Hamiltonian
using replica analysis and the absolute temperature zero limit\cite{Ciliberti2007}.
Shinzato et al.\ proposed a 
resolving algorithm for optimal portfolios based on the belief propagation method. This algorithm does not require the inverse Wishart matrix defined by return rate and those authors validated the efficacy of this algorithm via numerical experiments\cite{10.1371/journal.pone.0134968}.
Kondor et al.\ discuss the mean-variance model, the opportunity loss that can be defined by the ratio of the minimum expected investment risk of the annealed disordered system to the minimum investment risk of the quenched disordered system, and the distribution of opportunity loss. Using numerical experiments, those authors showed that there is a phase transition\cite{KONDOR20071545}.
Pafka et al.\ carried out stochastic optimization in a situation where the probability distribution of return rate was known; they evaluated the investment risk corresponding to learning errors and generalized errors in statistical learning theory using numerical experiments\cite{Pafka2002,PAFKA2003487}.
Using replica analysis, Ciliberti et al.\ investigated the optimal portfolio of the investment risk minimization problem of the expected shortfall model when the distribution of return rate of stock in each term was unknown, and clarified the phase diagram\cite{doi:10.1080/14697680701422089}.
Caccioli et al.\ discussed the instability of the cost function of the expected shortfall model {normalized}  by the L2 norm using replica analysis\cite{doi:10.1080/1351847X.2011.601661}.
Shinzato established short selling restrictions on the mean-variance model, {derived} the minimum investment risk and the optimal solution using replica analysis and the belief propagation method and showed that the minimum investment risk has a cusp region when the investment period ratio is at one half\cite{110008689817}.
Kondor et al.\ also set short selling restrictions for the mean-variance model when the expected return rate of each issue is unknown and evaluated the minimum investment risk using replica analysis\cite{1742-5468-2017-12-123402}.
Varga-Haszonits et al.\ discussed the stability of the opportunity loss and the replica symmetric solution for the investment risk minimization problem in which budget and expected return constraints were imposed but the distribution of return rate in each term was unknown\cite{1742-5468-2016-12-123404}.
As a first step towards evaluating the utility function, Shinzato used replica analysis to evaluate the optimal solution of the investment risk minimization problem with constraints in terms of investment cost, expected return, and budget. It was confirmed that the results accord with those obtained by the Lagrange undetermined multiplier method\cite{doi:10.7566/JPSJ.87.064801}.
Shinzato also used replica analysis to decentralize investments in multiple projects and discussed the net present value maximization problem imposed by concentrated investment and budget constraints to confirm that the internal return rate differs between the quenched disordered system and the annealed disordered system\cite{2018arXiv181006366S}.

Various arguments have been put forward in the literature for the mean-variance model. Shinzato used the Chernoff inequality and replica analysis to show that the minimum investment risk of the budget-constrained investment risk minimization problem satisfies the self-averaging property
\cite{10.1371/journal.pone.0133846}.
Moreover, Shinzato used replica analysis to explore the investment risk minimization problem with budget constraints when the distribution of return rate is distinct for each asset and assessed the optimal solution derived by the belief propagation method\cite{PhysRevE.94.062102b}.
Shinzato used replica analysis to evaluate the investment risk minimization problem with budget and concentrated investment constraints, as well as the concentrated investment minimization problem and its maximization problem, with budget and investment risk constraints, and went on to reveal a primal-dual relationship with respect to the results obtained in the previous research\cite{1742-5468-2017-2-023301,SHINZATO2018986}.
Tada et al.\ used the asymptotic eigenvalue distribution of the random matrix to analyze the investment risk minimization problem with budget and concentrated investment constraints, and their findings accord with results derived using replica analysis\cite{1742-5468-2017-2-023301,SHINZATO2018986,doi:10.7566/JPSJ.86.124804}.
Using the asymptotic eigenvalue distribution of the random matrix, Wakai et al.\ analyzed the budget-constrained investment risk minimization problem when the return rate of each asset does not follow a normal distribution and they clarified the relationship between the minimum investment risk and the variance of the return rate\cite{Ryosuke-Wakai2014}.
Based on that, Shinzato used replica analysis to evaluate the minimum investment risk of the mean-variance model when the rate of return is described by a one-factor model, {and to discuss} the relationship between the common factor and the minimum investment risk\cite{doi:10.7566/JPSJ.86.063802}.
Shinzato also examined the investment risk minimization problem with budget and expected return constraints as the primal problem, and the expected return maximization problem with budget and investment risk constraints as the dual problem. This was accomplished using the Lagrange undetermined multiplier method and replica analysis. It transpired that the optimal portfolio with respect to the minimum investment risk corresponds to the portfolio with the maximum expected return\cite{PhysRevE.94.052307}.
Moreover, Shinzato clarified that the dual structure is satisfied even in the quenched disordered system, and succeeded in constructing the macroscopic theory such as the opportunity loss which holds regardless of the distribution of the return rate, and the Pythagorean theorem of the Sharpe ratio\cite{1742-5468-2018-2-023401}.

However, in the extant literature, insufficient attention has been paid to the relationship between diversified investments and risk-free assets represented by deposits, savings, pensions, and government bonds.
Risk-free assets are important in the context of Tobin's separation theorem and the capital asset pricing model. Such assets are important to {develop financial theory,} and there is a pressing need to robustly evaluate the influence of a risk-free asset on the solution to the portfolio optimization problem. {The existing literature goes some way towards setting the necessary context for such work in terms of salient constraints \cite{1742-5468-2018-2-023401}, but} it remains necessary to extend the modeling approach to incorporate a risk-free asset.
{Accordingly, in this research, we improve the already discussed method for the case including a risk-free asset and analyze the investment risk minimization problem with budget and expected return constraints, which have been discussed in prior research.} However, we depart from the existing literature by using a cumulant generating function to discuss this problem in the context where a risk-free asset is available for inclusion in the portfolio alongside risky assets.
Moreover, we also discuss in detail {a macroscopic} theory of the problem such as the opportunity loss obtained in previous research\cite{1742-5468-2018-2-023401}, as well as the Pythagorean theorem of the Sharpe ratio and Tobin's separation theorem, which to the best of our knowledge have not hitherto been discussed in the literature.

The remainder of the {present} paper is organized as follows. In section \ref{sec2}, the investment risk minimization problem with budget and expected return constraints is extended. Next, in section \ref{sec3}, we use the Lagrange undetermined multiplier method to reformulate the portfolio optimization problem and derive the optimal solution and the minimum investment risk. Furthermore, because it is difficult to directly evaluate three moments, a novel cumulant generating function is defined in section \ref{sec4}, and the second derivative of the cumulant generating function is used to obtain those moments for the purpose of evaluating the minimum investment risk.
Section \ref{sec5} discusses some considerations to the analytical solution obtained by the proposed method, and section \ref{sec6} confirms the validity of the proposed method using numerical experiments. Finally, concluding remarks and suggestions for future research are put forward in {section \ref{sec7}. }

\section{Mean variance model\label{sec2}}
We consider the situation where investment occurs with respect to $N$ assets for $p$ periods in a stationary stock market without short-selling restrictions.
The asset portfolio $i(=1,2,\cdots,N-1)$ describes $w_i\in{\bf R}$ and
it is assumed that asset $N$ is risk-free, and its portfolio describes $w_N=N\rho$, where 
$\rho\in{\bf R}$ denotes the investment ratio of the risk-free asset.
Moreover, the return rate of asset $i$ at period $\mu$, $\bar{x}_{i\mu}$, is independently distributed with mean $E_X[\bar{x}_{i\mu}]=r_i$ and 
variance $V_X[\bar{x}_{i\mu}]=v_i$, and the return rate of 
asset $N$ at period $\mu$, $\bar{x}_{N\mu}$, is independently distributed with mean $E_X[\bar{x}_{N\mu}]=R_0$ and variance 
$V_X[\bar{x}_{N\mu}]=0$.
Following previous work\cite{1742-5468-2018-2-023401},
the budget constraint $\sum_{i=1}^Nw_i=N$ and expected return constraint $\sum_{i=1}^Nr_iw_i=NR$ are
rewritten using the portfolio of assets 1 to $N-1$
$\vec{w}=(w_1,\cdots,w_{N-1})^{\rm T}\in{\bf R}^{N-1}$,
\bea
\label{eq1}\sum_{i=1}^{N-1}w_i\eq N-N\rho,\\
\label{eq2}\sum_{i=1}^{N-1}r_iw_i\eq NR-N\rho R_0,
\eea
where $R\in{\bf R}$ is the coefficient which can characterize expected return and the notation
${\rm T}$ denotes the transpose of a vector or matrix. Thus, 
the investment risk ${\cal H}(\vec{w}|X)$ 
of portfolio $\vec{w}$ in the mean-variance model is 
defined by the sum of 
the squared differences between the whole return rate at each period $\sum_{i=1}^{N-1}w_i\bar{x}_{i\mu}$
and its mean $\sum_{i=1}^{N-1}w_ir_i$,
\bea
\label{eq3}
{\cal H}(\vec{w}|X)
\eq
\f{1}{2N}\sum_{\mu=1}^p\left(\sum_{i=1}^{N-1}w_i\bar{x}_{i\mu}-\sum_{i=1}^{N-1}w_ir_i\right)^2
\nn
\eq
\f{1}{2}\vec{w}^{\rm T}J\vec{w},
\eea
where the modified return rate $x_{i\mu}=\bar{x}_{i\mu}-r_i$ is used and 
similar to Hebb's law (e.g., the Hopfield model) the 
$i,j$th element of Wishart matrix $J=\left\{J_{ij}\right\}\in{\bf R}^{(N-1)\times(N-1)}$,  $J_{ij}$, is 
\bea
J_{ij}
\eq\f{1}{N}
\sum_{\mu=1}^px_{i\mu}x_{j\mu}.
\eea
It is noted that the modified return rate $x_{i\mu}$ is 
$E_X[x_{i\mu}]=0$ and $V_X[x_{i\mu}]=v_i$,  and 
from the result in \siki{eq7}, $p>N$ is needed.

From this, 
the optimal portfolio of the investment risk minimization problem with a risk-free asset $\vec{w}^*$ is 
\bea
\label{eq5}
\vec{w}^*\eq
\arg\mathop{\min}_{\vec{w}\in{\cal W}}
{\cal H}(\vec{w}|X),
\eea
where in \siki{eq5}
the feasible portfolio subset space which can satisfy
the budget constraint in \siki{eq1}
and the expected return constraint in 
\siki{eq2} of $\vec{w}\in{\bf R}^{N-1}$ is
\bea
{\cal W}=\left\{\vec{w}|\vec{w}^{\rm T}\vec{e}=N(1-\rho),\vec{w}^{\rm T}\vec{r}=N(R-\rho R_0)\right\},
\eea
where 
$\vec{e}=(1,1,\cdots,1)^{\rm T}\in{\bf R}^{N-1}$ and 
$\vec{r}=(r_1,r_2,\cdots,r_{N-1})^{\rm T}\in{\bf R}^{N-1}$ are already used.

\section{Lagrange undetermined multiplier method
\label{sec3}}
Here, let us analyze the investment risk minimization problem with a risk-free asset 
using the Lagrange undetermined multiplier method.
First, the Lagrange multiplier function is defined as 
\bea
L\eq
{\cal H}(\vec{w}|X)
+k\left(N(1-\rho)-\vec{w}^{\rm T}\vec{e}\right)\nn
&&+\theta\left(N(R-\rho R_0)-\vec{w}^{\rm T}\vec{r}\right),
\eea
where $k,\theta$ are parameters related to 
the budget constraint in \siki{eq1} and 
the expected return constraint in \siki{eq2}.
Moreover, from the extremum of the Lagrange multiplier function, 
$\pp{L}{w_i}=\pp{L}{k}=\pp{L}{\theta}=0$,
\bea
\label{eq7}
\vec{w}^*
\eq k^*J^{-1}\vec{e}+\theta^* J^{-1}\vec{r},\\
\begin{pmatrix}
1-\rho\\
R-\rho R_0
\end{pmatrix}
\eq
\begin{pmatrix}
g(0)&g(1)\\
g(1)&g(2)
\end{pmatrix}
\begin{pmatrix}
k^*\\
\theta^*
\end{pmatrix}
\label{eq8}
\eea
are obtained, where in \siki{eq8},
\bea
\label{eq9}
g(0)\eq\f{1}{N}\vec{e}^{\rm T}J^{-1}\vec{e},\\
g(1)\eq\f{1}{N}\vec{r}^{\rm T}J^{-1}\vec{e},\\
\label{eq11}
g(2)\eq\f{1}{N}\vec{r}^{\rm T}J^{-1}\vec{r}
\eea
are used. From this, 
\bea
\label{eq12}
k^*\eq
\f{1}{V_1g(0)}\left((1-\rho)\f{g(2)}{g(0)}-(R-\rho R_0)\f{g(1)}{g(0)}\right),\qquad\\
\label{eq13}
\theta^*\eq
\f{1}{V_1g(0)}\left(
R-\rho R_0-
(1-\rho)\f{g(1)}{g(0)}\right)
\eea
are obtained, where 
\bea
\label{eq14}
V_1\eq
\f{g(2)}{g(0)}
-\left(\f{g(1)}{g(0)}\right)^2
\eea
is set. From this, 
the minimal investment risk per asset $\ve=\f{1}{N-1}{\cal H}(\vec{w}^*|X)=\f{N}{N-1}\f{k^*(1-\rho)+\theta^*(R-\rho R_0)}{2}$ is, in the limit of 
the number of assets $N$, summarized as
\bea
\label{eq15}
\ve=
\f{1}{2g(0)}\left\{(1-\rho)^2+\f{\left(R-\rho R_0-(1-\rho)\f{g(1)}{g(0)}\right)^2}{V_1}\right\}.\eea

Based on this, 
if we can robustly evaluate the moments and parameters from \siki{eq9} to 
\siki{eq14}, 
the minimal investment risk per asset 
$\ve$ is rigorously assessed in
\siki{eq15}. However, 
when the moments from \siki{eq9}
to \siki{eq11} are estimated, 
we need to calculate the inverse Wishart matrix $J$, $J^{-1}$. 
Since the computational complexity here is $O(N^3)$, it is well known that 
it is difficult to evaluate the inverse matrix when the number of assets 
$N$ is large. Thus, 
hereafter, as an alternative approach, 
we accept the logarithmic function of the moment generating function, that is, the cumulant generating function to assess $g(0),g(1),g(2)$.

\section{Cumulant generating function and replica analysis
\label{sec4}}
Here we discuss 
evaluation of $g(0),g(1),g(2)$
without the inverse matrix $J^{-1}$.
First, the moment generating function is 
defined as 
\bea
Z\eq
\area 
\f{d\vec{w}}{(2\pi)^{\f{N-1}{2}}}
e^{
-\f{1}{2}\vec{w}^{\rm T}J\vec{w}
+k\vec{w}^{\rm T}\vec{e}
+\theta\vec{w}^{\rm T}\vec{r}
},
\eea
where the constant term with respect to the derivation of $k,\theta$ is ignored.
It is straightforward to analyze the moment generating function,  
\bea
\log Z
\eq-\f{1}{2}
\log\det
\left|J\right|+\f{k^2}{2}\vec{e}^{\rm T}J^{-1}\vec{e}+
k\theta\vec{r}^{\rm T}J^{-1}\vec{e}\nn
&&+
\f{\theta^2}{2}\vec{r}^{\rm T}J^{-1}\vec{r}.
\eea
From this, the logarithmic function of the moment generating function per asset, that is, the cumulant generating function $\phi=\lim_{N\to\infty}
\f{1}{N-1}\log Z$ is estimated as 
\bea
\phi
\eq-\f{1}{2}\lim_{N\to\infty}\f{1}{N-1}\log \det|J|\nn
&&
+\f{k^2}{2}g(0)+k\theta g(1)+\f{\theta^2}{2}g(2).
\eea
It transpires that $g(0),g(1),g(2)$ {are estimated by} the deviation of $\phi$ with respect to $k,\theta$. Thus, to evaluate 
$\phi$, we employ replica analysis. When $n\in{\bf Z}$, $E_X[Z^n]$ is summarized as 
\bea
&&\log E_X[Z^n]\nn
\eq-\f{p}{2}\log\det|I+Q_s|
+\f{N-1}{2}{\rm Tr}Q_s\tilde{Q}_s
-\f{n}{2}\sum_{i=1}^{N-1}\log v_i
\nn
&&-\f{N-1}{2}\log\det|\tilde{Q}_s|
+\f{1}{2}\vec{e}^{\rm T}\tilde{Q}_s^{-1}\vec{e}
\sum_{i=1}^{N-1}\f{(k+r_i\theta)^2}{v_i},\nn
\eea
where the identity matrix $I\in{\bf R}^{n\times n}$ is used and $E_X[f(X)]$ denotes  
the configuration average of $f(X)$ with respect to the return rate matrix $X=\left\{\f{x_{i\mu}}{\sqrt{N}}\right\}\in{\bf R}^{(N-1)\times p}$.
Moreover, the matrix of order parameters $Q_s=\left\{q_{sab}\right\}\in{\bf R}^{n\times n}$ and the matrix of these auxiliary order parameters $\tilde{Q}_s=\left\{\tilde{q}_{sab}\right\}\in{\bf R}^{n\times n}$, 
constant vector $\vec{e}=(1,1,\cdots,1)^{\rm T}\in{\bf R}^n$ are already accepted.
From this, in the limit of the number of assets 
$N$, it is summarized as
\bea
\label{eq20}
\psi(n)\eq
\lim_{N\to\infty}\f{1}{N-1}\log E_X[Z^n]\nn
\eq
\mathop{\rm Extr}_{Q_s,\tilde{Q}_s}
\left\{
-\f{\a}{2}\log\det|I+Q_s|
+\f{1}{2}{\rm Tr}Q_s\tilde{Q}_s\right.\nn
&&-\f{n}{2}
\kitai{\log v}-\f{1}{2}\log\det|\tilde{Q}_s|\nn
&&\left.
+\f{1}{2}
\vec{e}^{\rm T}\tilde{Q}_s^{-1}\vec{e}
\kitai{\f{(k+r\theta)^2}{v}}\right\},
\eea
where 
the investment period ratio 
$\a=p/(N-1)\sim O(1)$ is used. Further,
the notation $\mathop{\rm Extr}_uf(u)$
denotes the extremum of $f(u)$ with respect to parameter $u$ and the notation 
\bea
\kitai{f(r,v)}
\eq\lim_{N\to\infty}
\f{1}{N-1}
\sum_{i=1}^{N-1}f(r_i,v_i)
\eea
is used. From the extremum condition of 
\siki{eq20},
\bea
\label{eq22}
Q_s\eq\f{1}{\a-1}I+\f{\a}{(\a-1)^3}
\kitai{\f{(k+r\theta)^2}{v}}D,\\
\label{eq23}
\tilde{Q}_s\eq(\a-1)I-\f{1}{\a-1}
\kitai{\f{(k+r\theta)^2}{v}}
D
\eea
are obtained, where the constant matrix with whole element 1, $D\in{\bf R}^{n\times n}$, is used. We ignore $O(n)$ in the following evaluation of $\phi$. Moreover, when evaluating
\siki{eq20}, 
{it is noted that 
the ansatz of the replica symmetry solution 
is not assumed and we can calculate  
\sikis{eq22}{eq23}.}

From this, using 
$\phi=\lim_{n\to0}\pp{\psi(n)}{n}$,
\bea
\phi
\eq-\f{\a}{2}\log\f{\a}{\a-1}-\f{1}{2}\log(\a-1)-\f{1}{2}
\kitai{\log v}\nn
&&+\f{1}{2(\a-1)}
\kitai{\f{(k+r\theta)^2}{v}}
\eea
is estimated. Then, 
using $g(0)=\pp{^2\phi}{k^2}$, $
g(1)=\pp{^2\phi}{k\p \theta}
$ and $
g(2)=\pp{^2\phi}{\theta^2}
$,
\bea
g(0)\eq\f{\kitai{v^{-1}}}{\a-1},\\
g(1)\eq\f{\kitai{v^{-1}r}}{\a-1},\\
g(2)\eq\f{\kitai{v^{-1}r^2}}{\a-1}
\eea
are obtained. Substituting them into 
\siki{eq15},
\bea
\ve=
\f{\a-1}{2\kitai{v^{-1}}}
\left(
(1-\rho)^2
+
\f{\left(R-\rho R_0-(1-\rho)R_1\right)^2}{V_1}
\right)
\label{eq28}
\eea
is assessed, where 
\bea
R_1\eq\f{\kitai{v^{-1}r}}{\kitai{v^{-1}}},\\
V_1\eq
\f{\kitai{v^{-1}r^2}}{\kitai{v^{-1}}}
-\left(\f{\kitai{v^{-1}r}}{\kitai{v^{-1}}}\right)^2
\eea
are already used.

$R_1$ is the weighted average of the return rate of risky assets and $V_1$ is the weighted variance of the return rate of risky assets. 
Moreover, when $R_1<R_0$, the average of the return rate of risky assets is lower than the return rate of the risk-free asset. That is, 
since the optimal portfolio is a trivial investment strategy whereby rational investors prefer a zero-risk, high-return asset rather than high-risk, low-return assets, 
herein, mainly the case of $R_1\ge R_0$ is discussed.

\section{Discussion\label{sec5}}

\subsection{Minimal of minimal investment risk}
Although we succeeded  
in analytically deriving the minimal investment risk per asset in \siki{eq28} $\ve$, 
we did not sufficiently explore how the minimal investment risk behaves with respect to 
the investment ratio of the risk-free asset, that is, $\rho$. Thus, here let us solve the optimal $\rho=\rho^*$ which can minimize the minimal investment risk $\ve$.
From $\pp{\ve}{\rho}=0$,
for any $\a>1$,
\bea
\label{eq31}
\rho^*\eq
\f{V_1+(R-R_1)(R_0-R_1)}{V_1+(R_1-R_0)^2}
\eea
is obtained and substituted into \siki{eq28}. Then,
\bea
\label{eq32}
\ve_{\min}\eq
\f{\a-1}{2\kitai{v^{-1}}}
\f{(R-R_0)^2}{V_1+(R_1-R_0)^2}
\eea
is derived.
From \siki{eq31}, we can analytically determine  
the investment ratio of risk-free asset $w_N=N\rho$ which can minimize the minimal investment risk. The portfolio of risky assets is discussed in subsection \ref{sec5.7}.

\subsection{$R$ and $\rho^*$}
Next, we discuss the relationship between $R$ and $\rho^*$. From \siki{eq31},
\bea
\label{eq33}
\rho^*\le0
&\Longleftrightarrow&
R_1
+\f{V_1}{R_1-R_0}\le R,
\\
0<\rho^*<1
&\Longleftrightarrow&
R_0<R<R_1+\f{V_1}{R_1-R_0},\\
\rho^*\ge1
&\Longleftrightarrow&
R\le R_0
\label{eq35}
\eea
are obtained. From \siki{eq33},
when the expected return $R$ is large,
it means that rational investors borrow funds from risk-free asset $(\rho^*<0)$ and invest in risky assets.
From \siki{eq35},
when the expected return $R$ is small,
it also means that 
rational investors borrow
funds from risky asset $(\rho^*>1)$ and invest 
in the risk-free asset.

\subsection{Contextualizing the results}
Here, {let us compare our findings with the results obtained in cognate study \cite{1742-5468-2018-2-023401}, which discusses 
the investment risk minimization problem without a risk-free asset and the minimal investment risk per asset is obtained as follows:}
\bea
\label{eq36}
\ve_0\eq\f{\a-1}{2\kitai{v^{-1}}}
\left\{1+\f{(R-R_1)^2}{V_1}\right\}.
\eea
From this, we can compare $\ve_0$ with 
$\ve_{\min}$ in \siki{eq32}. Then,
\bea
\label{eq38}
\ve_0\ge \ve_{\min}
\eea
is obtained. That is, 
when investment stocks including a risk-free asset are compared with investment stocks that only contain risky assets, the latter can reduce investment risk compared to the former. Note that 
the equality of \siki{eq38} is at  
$R=R_1+\f{V_1}{R_1-R_0}$.

\subsection{Opportunity loss}
In this subsection, we compare 
the minimal expected investment risk discussed widely in operations research with the minimal investment risk derived in this paper.
Since the expected investment risk $E_X[{\cal H}(\vec{w}|X)]$
means the Hamiltonian of the annealed disordered system 
in spin glass theory,
\bea
E_X[{\cal H}(\vec{w}|X)]
\eq\f{\a}{2}\sum_{i=1}^{N-1}v_iw_i^2
\eea
is calculated. In this setting, also using the 
Lagrange undetermined multiplier method, 
the portfolio which can minimize 
the expected investment risk 
$E_X[{\cal H}(\vec{w}|X)]$, that is, 
$\vec{w}_{\rm OR}=\arg
\mathop{\min}_{\vec{w}\in{\cal W}}
E_X[{\cal H}(\vec{w}|X)]$, is straightforward to solve.
Thus, the Lagrange undetermined multiplier function is defined as follows:
\bea
L_{\rm OR}
\eq
E_X[{\cal H}(\vec{w}|X)]
+k(N(1-\rho)-\vec{w}^{\rm T}\vec{e})\nn
&&+\theta(N(R-\rho R_0)-\vec{w}^{\rm T}\vec{r}).
\eea
From the extremum conditions of $L_{\rm OR}$, that is, $\pp{L_{\rm OR}}{w_i}=
\pp{L_{\rm OR}}{k}=
\pp{L_{\rm OR}}{\theta}=0
$,
the minimal expected investment risk per asset 
$\ve_{\rm OR}=
\lim_{N\to\infty}\f{1}{N-1}\mathop{\min}_{\vec{w}\in{\cal W}}
E_X[{\cal H}(\vec{w}|X)]$ is assessed as 
\bea
\label{eq41-1}
\ve_{\rm OR}
=\f{\a}{2\kitai{v^{-1}}}
\left(
(1-\rho)^2
+
\f{\left(R-\rho R_0-(1-\rho)R_1\right)^2}{V_1}
\right).
\eea
Comparing $\ve$ in \siki{eq28} and $\ve_{\rm OR}$ in \siki{eq41-1},
the opportunity loss which is defined by the ratio of the minimal expected investment risk $\ve^{\rm OR}$ with respect to 
the minimal investment risk $\ve$,
that is, $\kappa=\f{\ve_{\rm OR}}{\ve}$, is estimated as
\bea
\label{eq42-1}
\kappa\eq\f{\a}{\a-1}.
\eea
From \siki{eq42-1},
since the opportunity loss $\kappa$ is only a function of the investment period ratio $\a$
and does not depend on the probabilities of $r_i,v_i$, 
the expected return rate of risk-free asset $R_0$ and 
the portfolio of risk-free assets $\rho$, 
it transpires that 
this macroscopic relationship always holds.
Moreover, as $\a$ converges towards 1, the opportunity loss 
becomes large,  
since $\vec{w}_{\rm OR}
=\arg
\mathop{\min}_{\vec{w}\in{\cal W}}
E_X[{\cal H}(\vec{w}|X)]
$ is not consistent with 
$\vec{w}^*$ in \siki{eq7}.
Thus, the portfolio which is derived by the method of operations research, 
$\vec{w}_{\rm OR}$, is not appropriate for 
the optimal diversification of investments.

\subsection{Pythagorean theorem of the Sharpe ratio}
Let us discuss the macroscopic relation of the Sharpe ratio.
The Sharpe ratio is a criterion which is defined by 
the return per risk and is written, using the notation of Eqs. (\ref{eq2}), (\ref{eq3}), and (\ref{eq28}), as
\bea
\label{eq42}
S(R)\eq\f{R-\rho R_0}{\sqrt{2\ve}}.
\eea
It transpires that, 
from 
\siki{eq2}, 
the numerator 
$R-\rho R_0$ means the expected return rate of risky assets per asset and from \sikis{eq3}{eq28} 
the denominator $\sqrt{2\ve}$ means 
the standard deviation of the return rate of risky assets per asset. Then 
the expected return rate $R$ which can maximize the Sharpe ratio is derived as
\bea
\label{eq43}
R^*\eq\arg\mathop{\max}_RS(R)\nn
\eq
\rho R_0+(1-\rho)\left(R_1+\f{V_1}{R_1}\right),
\eea
and the square of the maximum of the Sharpe ratio $S^2(R^*)$ is estimated as
\bea
\label{eq44}
S^2(R^*)\eq
\f{V_1+R_1^2}{\a-1}
\kitai{v^{-1}}.
\eea
Furthermore, two of the expected return rates $R$
which can minimize and maximize 
$\ve=\ve(R)$ in \siki{eq28} are respectively
assessed as
\bea
R_{\min}\eq\arg\mathop{\min}_R\ve(R)\nn
\eq\rho R_0+(1-\rho)R_1,\\
R_{\max}\eq\arg\mathop{\max}_R\ve(R)\nn
\eq\infty,
\eea
and the squares of the Sharpe ratio are evaluated as
\bea
S^2(R_{\min})\eq
\f{R_1^2}{\a-1}
\kitai{v^{-1}},\\
S^2(R_{\max})\eq
\f{V_1}{\a-1}
\kitai{v^{-1}}.
\eea
Then, the macroscopic relation
\bea
S^2(R^*)\eq
S^2(R_{\min})
+S^2(R_{\max})
\eea
is obtained. {Similar to in \siki{eq42-1},} 
it transpires that 
the Pythagorean theorem of the Sharpe ratio also holds in the case of a risk-free asset and risky assets. Moreover, it is shown that this Pythagorean theorem 
does not depend on 
the probabilities of 
$r_i,v_i$, the mean of the return rate of the risk-free asset 
$R_0$, the portfolio of risk-free asset $\rho$, and 
the investment period ratio $\a$.

\subsection{Maximal Sharpe ratio}
Here using the Cauchy--Schwarz inequality 
$(\vec{a}^{\rm T}\vec{b})^2\le{\vec{a}^{\rm T}\vec{a}\cdot\vec{b}^{\rm T}\vec{b}}$,
we discuss the maximal Sharpe ratio.
From Eqs. (\ref{eq2}), (\ref{eq3}), and (\ref{eq28}), the Sharpe ratio in \siki{eq42}
is replaced by
\bea
S(R)\eq\f{\f{1}{N}\vec{w}^{\rm T}\vec{r}}
{\sqrt{\f{1}{N-1}\vec{w}^{\rm T}J\vec{w}}}.
\eea
Further, setting $\vec{a}=J^{\f{1}{2}}\vec{w}$ and $\vec{b}=J^{-\f{1}{2}}\vec{r}$, since 
The Cauchy--Schwarz inequality is calculated as 
$(\vec{a}^{\rm T}\vec{b}/\sqrt{\vec{a}^{\rm T}\vec{a}})^2\le{\vec{b}^{\rm T}\vec{b}}$,
in the limit of a large number of assets $N$,
\bea
\label{eq52}
S^2(R)&\le&{\f{N-1}{N}}
\f{\vec{r}^{\rm T}J^{-1}\vec{r}}{N}\nn
\eq{\f{\kitai{v^{-1}r^2}}{\a-1}}
\eea
is obtained. From $\kitai{v^{-1}r^2}=\kitai{v^{-1}}(V_1+R_1^2)$,
it is indicated that the right-hand side in \siki{eq52}
is consistent with \siki{eq44}. In addition, 
from the equal condition of this inequality 
$\vec{a}=l\vec{b}$
($l$ is a scalar coefficient),
$\vec{w}=lJ^{-1}\vec{r}$ is obtained. Comparing this and \siki{eq7}, it transpires that $k^*=0$ also holds. That is, using \siki{eq12} from $k^*=0$,
\bea
R^*\eq
\rho R_0+(1-\rho)\left(R_1+\f{V_1}{R_1}\right)
\eea
is also obtained and is consistent with \siki{eq43}.

\subsection{{Tobin's separation} theorem\label{sec5.7}}
Finally, let us discuss
{Tobin's separation} theorem for investment stocks including a risk-free asset. 
Using a finding obtained in previous work (that is, \siki{eq36}), 
the standard deviation of the return rate of risky assets 
$y(R)=\sqrt{2\ve_0}$ is estimated as follows:
\bea
y(R)\eq\sqrt{\f{\a-1}{\kitai{v^{-1}}}
\left(1+\f{(R-R_1)^2}{V_1}\right)
}.
\eea
Moreover, 
from the mean of the return rate of risk-free asset $R_0$ and
 its standard deviation, that is, 0, we can easily assess the 
tangent line from {the point $(R_0,0)$ to }
the function $y(R)$. 
The coordinate of tangent point M is set as
$(R_M,y(R_M))$. We refer to  
the portfolio at tangent point M as the market portfolio.
From this, 
$R_M$ and $y(R_M)$ are 
derived {using the condition that the 
slope at tangent point M is consistent with 
the slope of a straight line connecting $(R_0,0)$ and 
$(R_M,y(R_M))$. That is, they are solved from the relation 
$y'(R_M)=\f{y(R_M)-0}{R_M-R_0}$. Then, 
}
\bea
\label{eq55-1}
R_M\eq
R_1+\f{V_1}{R_1-R_0},\\
y(R_M)\eq
\sqrt{\f{\a-1}{\kitai{v^{-1}}}
\left(1+\f{V_1}{(R_1-R_0)^2}\right)
}
\eea
are obtained. Using the two-fund separation theorem, 
for an arbitrary expected return rate coefficient $R$,
{the $y$-coordinate of point A $(R,y_A(R))$ on} the tangent line 
$y_A(R)=\f{R-R_0}{R_M-R_0}y(R_M)+\f{R_M-R}{R_M-R_0}\cdot0$
is calculated as
\bea
\label{eq56}
y_A(R)
\eq\sqrt{\f{\a-1}{\kitai{v^{-1}}}}\f{R-R_0}{\sqrt{V_1+(R_1-R_0)^2}},
\eea
where it transpires that 
\siki{eq56} denotes the capital allocation line of the quenched disordered system.

Using $\ve_{\min}$ in \siki{eq32},
$y_{\min}(R)=\sqrt{2\ve_{\min}}$ is
\bea
y_{\min}(R)\eq
\sqrt{\f{\a-1}{\kitai{v^{-1}}}}\f{R-R_0}{\sqrt{V_1+(R_1-R_0)^2}};
\eea
since it is consistent with \siki{eq56}, it is determined that the portfolio which is composed of two assets, the representative asset and the risk-free asset, is consistent with the optimal portfolio which can minimize investment risk {in \siki{eq3}}. Further, 
note that 
$R_M$ in \siki{eq55-1} is consistent with 
the equality condition of \siki{eq38}.

Additionally, substituting 
$\rho^*$
in \siki{eq31} into $\rho$ of $k^*$ in \siki{eq12} and 
$\theta^*$ in \siki{eq13},
\bea
k^*\eq-\f{R-R_0}{g(0)(V_1+(R_1-R_0)^2)}R_0,\\
\theta^*\eq
\f{R-R_0}{g(0)(V_1+(R_1-R_0)^2)}
\eea
are obtained. We also substitute them into  $\vec{w}^*=k^*J^{-1}\vec{e}+\theta^*J^{-1}\vec{r}$ in \siki{eq7} and obtain
\bea
\label{eq60}
\vec{w}^*\eq
\f{R-R_0}{g(0)(V_1+(R_1-R_0)^2)}J^{-1}\left(\vec{r}-R_0\vec{e}\right).\qquad
\eea
From this finding in \siki{eq60},
although the portfolio of risk-free asset 
$w_N^*=N\rho^*$ depends on {the coefficient of the expected return rate $R$ from \siki{eq31},}
the investment ratio of each risky asset, that is, asset 1
 to asset $N-1$, does not depend on $R$.
Thus, for any $R$,
since 
the direction of the vector $\vec{w}^*$, 
$J^{-1}(\vec{r}-R_0\vec{e})$, does not change,  
it transpires that 
$w_i^*/w_j^*=const.(i,j=1,2\cdots,N-1)$ holds.
Note that 
$\vec{w}^*$ in \siki{eq60} is just the market portfolio 
and {Tobin's separation} theorem of the quenched disordered system is indicated.

\section{Numerical experiments\label{sec6}}
In this section, we use the results of numerical experiments to confirm the validity of 
the analytical results of the minimal investment risk per asset $\ve$ and the Sharpe ratio $S$ which are evaluated by 
$g(0),g(1),g(2)$
derived from the cumulant generating function $\phi$.
We take  
the mean of the return rate of asset $i$
$E_X[\bar{x}_{i\mu}]$ as $r_i$ and its variance $V_X[\bar{x}_{i\mu}]$ as $v_i=h_ir_i^2$;
that is, the variance is the product of the square of the average of return rate $r_i^2$ and 
the random proportionality coefficient 
$h_i(>0)$.
Furthermore, it is assumed that 
$r_i$ and $h_i$ are independently distributed with the following bounded Pareto distributions which are defined by $(l_r\leq r_i\leq u_r,l_h\leq h_i\leq u_h)$:
\bea
\label{eq4-41}
 f_r(r_i)=
\left\{\begin{array}{ll}
\f{1-c_r}{u_r^{1-c_r}-l_r^{1-c_r}}r_i^{-c_r} &l_r\leq r_i\leq u_r \\
      0&{\rm otherwise} \end{array}\right., \\
\label{eq4-42}
 f_h(h_i)=
\left\{\begin{array}{ll}
\f{1-c_h}{u_h^{1-c_h}-l_h^{1-c_h}}h_i^{-c_h} &l_h\leq r_i\leq u_h \\
      0&{\rm otherwise} \end{array}\right.,
\eea
where the parameters of the bounded Pareto distributions of $r_i,h_i$, $f_r(r_i),f_h(h_i)$,
are accepted as $(l_r,u_r,c_r)$ and $(l_h,u_h,c_h)$, respectively. 
{For convenience, $l_r,l_h,c_r,c_h>0$ are assumed to be satisfied.}

\begin{figure}[tb] 
\begin{center}
\includegraphics[width=1.0\hsize]{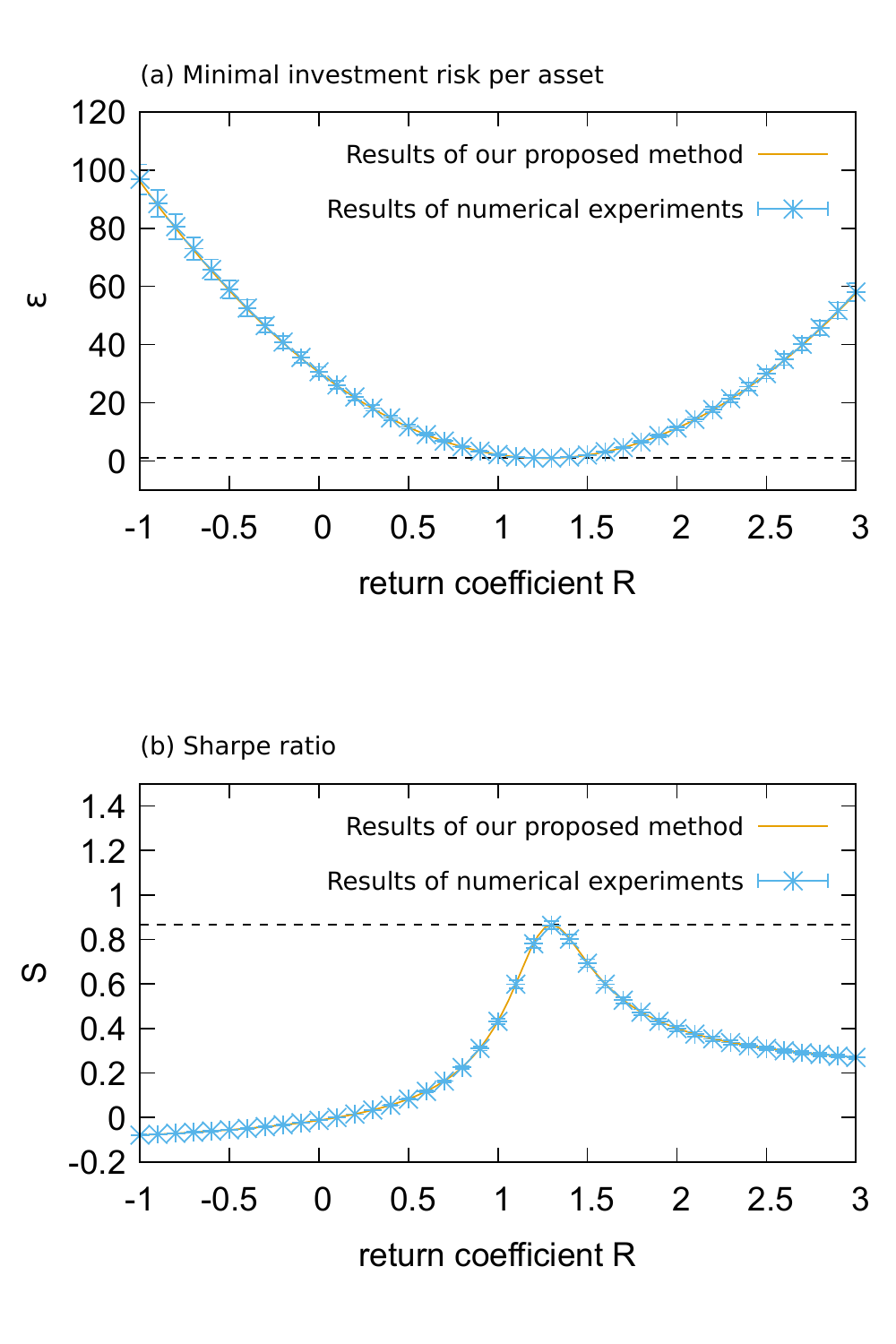}
\caption{
\label{Fig1}
Comparison of results generated from the proposed method and numerical experiments.
}
\end{center}
\end{figure}

Using the following procedure,
we numerically estimate the minimal investment risk per asset $\ve$ and Sharpe ratio $S$.
\begin{description}
\item[Step 1] 
$r_i,h_i$ are randomly assigned by \sikis{eq4-41}{eq4-42}, and {then}
$r_i,v_i(=h_ir_i^2)$ are prepared.
\item[Step 2]The return rate of asset $i$
$\bar{x}_{i\mu}$ is also randomly assigned by the distribution with $E_X[\bar{x}_{i\mu}]=r_i$ and $V_X[\bar{x}_{i\mu}]=v_i$. Furthermore,
the modified return rate $x_{i\mu}=\bar{x}_{i\mu}-r_i$ is calculated, and
the matrix of return rate $X=\left\{\f{x_{i\mu}}{\sqrt[]{\mathstrut N}}\right\}\in{\bf R}^{(N-1)\times p}$ is set.
\item[Step 3]We set $J=XX^{\rm T}\in{\bf R}^{(N-1)\times (N-1)}$ and calculate its inverse matrix $J^{-1}$.
\item[Step 4]$g(0)=\f{1}{N}\vec{e}^{\rm T}J^{-1}\vec{e}$, $g(1)=\f{1}{N}\vec{r}^{\rm T}J^{-1}\vec{e}$, $g(2)=\f{1}{N}\vec{r}^{\rm T}J^{-1}\vec{r}$ are assessed.
\item[Step 5]
Using $g(0),g(1),g(2)$,
\bea
V_1\eq
\f{g(2)}{g(0)}-\left(\f{g(1)}{g(0)}\right)^2,\\
k^*\eq
\f{1}{g(0)V_1}\left[(1-\rho)\f{g(2)}{g(0)}-(R-\rho R_0)\f{g(1)}{g(0)}\right],\nn
\\
\theta^*\eq
\f{1}{g(0)V_1}\left[R-\rho R_0-(1-\rho)\f{g(1)}{g(0)}\right]
\eea
are numerically obtained.
\item[Step 6]Using 
$\ve=\f{k(1-\rho)+\theta(R-\rho R_0)}{2}\times\f{N}{N-1}$ and $S=\f{R-\rho R_0}{\sqrt[]{\mathstrut 2\ve}}$, $\ve$ and $S$ are estimated.
\end{description}
In terms of experimental settings, we employ $N=1000,p=2000,(\a=2),l_r=l_h=1,u_r=u_h=c_r=c_h=2$ and $\rho=0.1,R_0=1$. We run 100 trials. From this, 
we can numerically estimate typical behaviors of the minimal investment risk per asset and the Sharpe ratio, and compare the results with {\sikis{eq28}{eq42}, shown in Fig. \ref{Fig1}.}
The horizontal and vertical axes 
in Fig. \ref{Fig1}(a) show the expected return rate $R$ and the minimal investment risk per asset $\ve$, respectively. 
The horizontal and vertical axes 
in Fig. \ref{Fig1}(b) show the expected return rate $R$ and the Sharpe ratio $S$, respectively.
The solid lines denote results generated by the proposed method and asterisks with error bars denote the numerical results. From both figures
it is clear that the results generated using the proposed method and the numerical experiment are consistent. In other words, the validity of the proposed method based on the cumulant generating function proposed herein is confirmed.

\section{Conclusions\label{sec7}}
We extended the analytical approach for solving the investment risk minimization problem with budget and expected return constraints. We solved the optimal portfolio which can minimize investment risk with these constraints, including a risk-free asset.
More specifically, one stock in the investment market is regarded as a risk-free asset, {and we explored the relationship between a portfolio comprising a risk-free asset along with risky assets and a portfolio comprising only risky assets.} The Lagrange undetermined multiplier method was used to reformulate the portfolio optimization problem. To analytically evaluate the optimal solution, although it was necessary to determine three moments so as to estimate the inverse of the Wishart matrix, as another approach, we derive the minimal investment risk using a method that can estimate three moments without directly determining the inverse matrix.
Through comparison with results from previous studies and through the use of numerical experiments, it is confirmed that a risk-free asset is important to minimize investment risk. Moreover, we succeeded in deriving the relation of opportunity loss, the Pythagorean theorem of the Sharpe ratio, and Tobin's separation theorem in the quenched disordered system.

Herein, our focus was on the mean-variance model and as such there is ample scope for future research to extend the scope to other investment risk models for the purpose of developing the theory of risk management. For example, analysis of the relationship between minimal investment risk and the investment ratio of the risk-free asset using the absolute deviation model and the expected shortfall model would be important lines of inquiry.

\section*{Acknowledgements}
The authors are grateful for fruitful discussions with D. Tada. 
This work was partially supported by
Grants-in-Aid Nos. 15K20999, 17K01260, and 17K01249; Research Project of the Institute of Economic Research Foundation at Kyoto University; and Research Project
No. 4 of the Kampo Foundation.

\appendix
\section{Replica analysis}
Since the 
investment risk minimization problem including a risk-free asset with budget and return constraints is regarded as the ground state estimation problem in canonical ensembles, 
in this appendix,
we resolve this portfolio optimization problem 
using replica analysis. {The partition function $Z$ of the Boltzmann distribution of 
the Hamiltonian ${\cal H}(\vec{w}|X)$
in \siki{eq3} of the inverse temperature $\b$ is represented by}
\bea
Z\eq
\int_{\vec{w}\in{\cal W}}
d\vec{w}e^{-\b{\cal H}(\vec{w}|X)}.
\eea
In the limit of a large number of assets $N$ and based on the replica symmetry ansatz,
\bea
\phi
\eq
\lim_{N\to\infty}\f{1}{N-1}
E_X\left[\log Z\right]\nn
\eq
\mathop{\rm Extr}_{\Theta}
\left\{
-\f{1}{2}
\kitai{\log(\tilde{\chi}_w+v\tilde{\chi}_s)}
+\f{1}{2}
\kitai{\f{\tilde{q}_w+v\tilde{q}_s}{\tilde{\chi}_w+v\tilde{\chi}}_s}
\right.\nn
&&
+\f{1}{2}
\kitai{\f{(k+r\theta)^2}{\tilde{\chi}_w+v\tilde{\chi}}_s}
-\f{\a}{2}\log(1+\b\chi_s)\nn
&&-\f{\a\b q_s}{2(1+\b\chi_s)}
-k(1-\rho)
-\theta(R-\rho R_0)\nn
&&+\f{1}{2}(\chi_w+q_w)(\tilde{\chi}_w-\tilde{q}_w)+\f{1}{2}q_w\tilde{q}_w\nn
&&\left.
+\f{1}{2}(\chi_s+q_s)(\tilde{\chi}_s-\tilde{q}_s)+\f{1}{2}q_s\tilde{q}_s
\right\}
\eea
is obtained, where 
the order parameters $q_{wab}=\f{1}{N-1}\sum_{i=1}^{N-1}w_{ia}w_{ib}$ and 
$q_{sab}=\f{1}{N-1}\sum_{i=1}^{N-1}v_iw_{ia}w_{ib}$ and their auxiliary order parameters $\tilde{q}_{wab},\tilde{q}_{sab}$ are used.
{In addition, the replica symmetry solution is set as}
\bea
q_{wab}\eq
\left\{
\begin{array}{ll}
\chi_w+q_w&a=b\\
q_w&a\ne b
\end{array}
\right.,\\
q_{sab}\eq
\left\{
\begin{array}{ll}
\chi_s+q_s&a=b\\
q_s&a\ne b
\end{array}
\right.,\\
\tilde{q}_{wab}\eq
\left\{
\begin{array}{ll}
\tilde{\chi}_w-\tilde{q}_w&a=b\\
-\tilde{q}_w&a\ne b
\end{array}
\right.,\\
\tilde{q}_{sab}\eq
\left\{
\begin{array}{ll}
\tilde{\chi}_s-\tilde{q}_s&a=b\\
-\tilde{q}_s&a\ne b
\end{array}
\right.,\\
k_a\eq k,\\
\theta_a\eq\theta,
\eea
where $(a,b=1,2,\cdots,n)$. 
Further, $k$ and $\theta$ are the parameters related to 
the budget constraint in \siki{eq1}
and 
the return constraint in \siki{eq2}.
The set of order parameters 
$\Theta=(k,\theta,\chi_w,q_w,\chi_s,q_s,\tilde{\chi}_w,\tilde{q}_w,\tilde{\chi}_s,\tilde{q}_s)$ is already applied.
From the extremum conditions of these order parameters, $\pp{\phi}{k}=\pp{\phi}{\theta}=\pp{\phi}{\chi_w}
=\pp{\phi}{q_w}
=\pp{\phi}{\chi_s}
=\pp{\phi}{q_s}
=\pp{\phi}{\tilde{\chi}_w}
=\pp{\phi}{\tilde{q}_w}
=\pp{\phi}{\tilde{\chi}_s}
=\pp{\phi}{\tilde{q}_s}
=0$,
\bea
k\eq\f{\b(\a-1)}{\kitai{v^{-1}}V_1}
\left((1-\rho)(V_1+R_1^2)
-(R-\rho R_0)R_1
\right),\nn
\\
\theta\eq\f{\b(\a-1)}{\kitai{v^{-1}}V_1}
\left(R-\rho R_0
-(1-\rho)R_1
\right),
\\
\chi_w\eq\f{\kitai{v^{-1}}}{\b(\a-1)},\\
q_w\eq
\f{1}{\a-1}
\left((1-\rho)^2+\f{(R-R_\rho)^2}{V_1}\right)\nn
&&+\f{\kitai{v^{-2}}}{\kitai{v^{-1}}^2}
\f{(1-\rho)^2V_2}{V_2+(R_2-R_1)^2}\nn
&&+
\f{\kitai{v^{-2}}}{\kitai{v^{-1}}^2}
\f{V_2+(R_2-R_1)^2}{V_1^2}\nn
&&\times
\left(R-R_\rho+\f{V_1(1-\rho)(R_2-R_1)}{V_2+(R_2-R_1)^2}\right)^2,
\\
\tilde{\chi}_w\eq0,\\
\tilde{q}_w\eq0,\\
\chi_s\eq\f{1}{\b(\a-1)},\\
q_s\eq
\f{\a}{(\a-1)\kitai{v^{-1}}}
\left((1-\rho)^2+\f{(R-R_\rho)^2}{V_1}\right),\qquad\quad\\
\tilde{\chi}_s\eq\b(\a-1),\\
\tilde{q}_s\eq
\f{\b^2(\a-1)}{\kitai{v^{-1}}}
\left((1-\rho)^2+\f{(R-R_\rho)^2}{V_1}\right)
\eea
are obtained analytically, where 
\bea
R_1\eq\f{\kitai{v^{-1}r}}{\kitai{v^{-1}}},\\
R_2\eq\f{\kitai{v^{-2}r}}{\kitai{v^{-2}}},\\
V_1\eq\f{\kitai{v^{-1}r^2}}{\kitai{v^{-1}}}
-\left(\f{\kitai{v^{-1}r}}{\kitai{v^{-1}}}\right)^2,\\
V_2\eq\f{\kitai{v^{-2}r^2}}{\kitai{v^{-2}}}
-\left(\f{\kitai{v^{-2}r}}{\kitai{v^{-2}}}\right)^2,\\
R_\rho\eq \rho R_0+(1-\rho)R_1
\eea
are used. From this, {since the minimal investment risk per asset $\ve$ is derived using the thermodynamic relation 
$\ve=-\lim_{\b\to\infty}\pp{\phi}{\b}$
and 
$-\pp{\phi}{\b}=\f{\a\chi_s}{2(1+\b\chi_s)}+\f{\a q_s}{2(1+\b\chi_s)^2}$, we infer that}
\bea
\ve\eq
\f{\a-1}{2\kitai{v^{-1}}}
\left(
(1-\rho)^2
+
\f{\left(R-R_\rho\right)^2}{V_1}
\right),
\eea
and it transpires that this is consistent with 
\siki{eq28}.

\section{Dual problem}
In this appendix, 
the dual problem of the main paper's primal problem is reformulated by the expected return maximization problem including a risk-free asset  
with budget and investment risk constraints, as follows:
\bea
\label{eq-b-1}
\mathop{\max}_{\vec{w}\in{\cal D}}
\left\{
\sum_{i=1}^Nr_iw_i
\right\},
\eea
where the feasible subset of the portfolio from asset 1 to asset $N-1$, $\vec{w}\in{\bf R}^{N-1}$, is 
\bea
{\cal D}
=
\left\{
\vec{w}
\in{\bf R}^{N-1}
\left|
\vec{w}^{\rm T}\vec{e}=N(1-\rho),
\f{1}{2}\vec{w}^{\rm T}J\vec{w}
=N\ve
\right.
\right\}.
\eea
In \siki{eq-b-1}, because of the 
optimization problem of risky assets, 
we do not optimize the portfolio of risk-free asset $w_N=N\rho$.

From this, using the Lagrange undetermined multiplier method, the Lagrange function 
is set as 
\bea
L\eq\sum_{i=1}^Nr_iw_i
+k\left(\sum_{i=1}^{N-1}w_i-N+N\rho\right)\nn
&&+\tau\left(N\ve-\f{1}{2}\vec{w}^{\rm T}J\vec{w}\right),
\eea
and from the extremum condition of the Lagrange function, $\pp{L}{w_i}=\pp{L}{k}=\pp{L}{\tau}=0$,
\bea
\tau^*\eq g(0)\sqrt{\f{V_1}{2\ve g(0)-(1-\rho)^2}},\\
k^*\eq(1-\rho)\sqrt{\f{V_1}{2\ve g(0)-(1-\rho)^2}}-R_1,\\
\vec{w}^*\eq\f{k^*}{\tau^*}J^{-1}\vec{e}+\f{1}{\tau^*}J^{-1}\vec{r}
\eea
are derived. Substituting them into 
\siki{eq-b-1}, the maximal expected return per asset $R$ is calculated as
\bea
R\eq\lim_{N\to\infty}\f{1}{N}\sum_{i=1}^{N-1}r_iw_i^*+\rho R_0\nn
\eq\rho R_0+(1-\rho)R_1+\f{g(0)}{\tau^*}V_1\nn
\eq\rho R_0+(1-\rho)R_1+
\sqrt{V_1(2\ve g(0)-(1-\rho)^2)},\nn
\eea
and it transpires that this is consistent with \siki{eq28}.

\bibliographystyle{jpsj}
\bibliography{sample20190105}

\end{document}